\documentclass[a4paper,10pt]{article}
\usepackage{amsmath}
\usepackage{amssymb}
\usepackage{graphicx}

\title{Strongly hyperbolic Hamiltonian systems in numerical relativity: Formulation and symplectic integration}

\author{Ronny Richter\\~\\
Mathematisches Institut, Universit\"at T\"ubingen,\\
Auf der Morgenstelle 10, 72076 T\"ubingen, Germany\\
richter@na.uni-tuebingen.de}

\newcommand\bfzero{{\mathbf 0}}

\newcommand\bfg{{\mathbf g}}

\newcommand\bfp{{\mathbf p}}
\newcommand\bfq{{\mathbf q}}

\newcommand\bfx{{\mathbf x}}

\newcommand\bfS{{\mathbf S}}
\newcommand\bfD{{\mathbf D}}
\newcommand\bfE{{\mathbf E}}

\newcommand\bfbeta{{\boldsymbol \beta}}
\newcommand\bfgamma{{\boldsymbol \gamma}}
\newcommand\bflambda{{\boldsymbol \lambda}}
\newcommand\bfmu{{\boldsymbol \mu}}

\newcommand\bfA{{\mathbf A}}

\newcommand\bfnabla{{\boldsymbol\nabla}}

\newcommand\dt{\Delta t}

\begin{document}

\maketitle

\begin{abstract}
We consider two strongly hyperbolic Hamiltonian formulations of general relativity and their numerical integration with a free and a partially constrained symplectic integrator. In those formulations we use hyperbolic drivers for the shift and in one case also for the densitized lapse. A system where the densitized lapse is an external field allows to enforce the momentum constraints in a holonomically constrained Hamiltonian system and to turn the Hamilton constraint function from a weak to a strong invariant.

These schemes are tested in a perturbed Minkowski and the Schwarz\-schild space-time. In those examples we find advantages of the strongly hyperbolic formulations over the ADM system presented in \cite{Richter:2008pr}. Furthermore we observe stabilizing effects of the partially constrained evolution in Schwarz\-schild space-time as long as the momentum constraints are enforced.
\end{abstract}

PACS numbers: 04.25.D-, 04.20.Fy
%\pacs{04.25.D-, 04.20.Fy}
%\submitto{\CQG}
%\maketitle

\section{Introduction}

Many of the recently developed schemes in numerical relativity are based on the BSSN system \cite{PhysRevD.52.5428,PhysRevD.59.024007}. One of the key features that distinguishes BSSN from e.g. the dynamical ADM equations is its strong hyperbolicity. Systems with that property are preferred, because  the well-posedness of the Cauchy problem can be proven \cite{Nagy:2004td} and often they lead to numerical schemes with improved stability properties \cite{calabrese_1228487}. Using strongly hyperbolic systems stable evolutions of the black hole binary problem are possible since 2005 \cite{pretorius-2005-95,campanelli-2006-96,baker-2006-96,Laguna-2006,bruegmann-2006}.

Here we are interested in numerical schemes that apply \emph{symplectic integrators} for Hamiltonian systems. These integration methods respect the Hamiltonian structure in the discretization and have been essential for attaining favorable propagation properties in various areas of scientific computing (see, e.g., \cite{HaLW06,LeR04,SSC94} and references). The problem there is that, in contrast to the dynamical ADM equations \cite{ADM_collection2191} which are just weakly hyperbolic, many strongly hyperbolic formulations known do not possess a Hamiltonian structure. Yet, there is one exception, namely the Hamiltonian \emph{generalized harmonic system} that Brown proposed in \cite{brown-2008}.

In this article we present a second strongly hyperbolic Hamiltonian system, the \emph{fixed lapse system}, and apply the free and constrained symplectic integrators introduced in \cite{Richter:2008pr} to these two systems. Other applications of symplectic integrators in numerical relativity are presented e.g. in
\cite{Berger_Garfinkle_CQG,Berger_Moncrief_PhysRevD.48.4676,
blanco_costa_1995,brown-2006-73,gambini-2005}.

We compare the properties of our schemes in numerical experiments with effectively 1+1-dimensional versions of Einstein's equations.
The purposes of these experiments are the following. At first we want to confirm the expectation that schemes based on the strongly hyperbolic systems lead to more stable results than schemes based on the (just weakly hyperbolic) ADM equations. Furthermore we want to recover the good propagation properties of nearly conserved quantities that we found in \cite{Richter:2008pr} with the new schemes here.
Finally, the major question addressed is, for which systems and in which situations constrained evolution is beneficial in comparison with free evolution. In \cite{Richter:2008pr} we found better results for constrained integration of the ADM equations, but it is not clear whether this is still valid when the free evolution schemes are based on strongly hyperbolic systems.

We start our considerations in section 2 where we introduce the two Hamiltonian formulations and discuss their pros and cons. Brown's Hamiltonian has the advantage that it deals with well-known gauge conditions, whereas for the fixed lapse system we can apply a numerical scheme that enforces the momentum constraints in a holonomically constrained Hamiltonian system.

In section 3 we describe the spatially discrete Hamiltonians and the symplectic integration methods used. In the free evolution schemes the integrator is the {\it St\"ormer--Verlet method}, a standard second order symplectic integrator (see \cite{HaLW03,HaLW06,LeR04}). There the Hamiltonian and momentum constraints are not taken into account and may drift off.

The problem of constraint growth is addressed by applying the {\it RATTLE method}, a constrained version of the St\"ormer-Verlet integrator. It is applicable to holonomically constrained Hamiltonian systems. For Brown's Hamiltonian it can be used to enforce a combination of the momentum constraints and a set of new constraints that arise since the densitized lapse becomes a dynamical variable. For the fixed lapse system we can achieve even more, namely that the momentum constraints themselves are enforced. The mechanism behind is the same as for the ADM equations \cite{Richter:2008pr}.

In section 4 we discuss the important aspects of the 1+1-dimensional implementation in detail and introduce the test examples with which we have done our numerical experiments: a perturbed Minkowski problem and Schwarzschild space-time in two different coordinate systems.

Section 5 compares four integration methods for a perturbed Minkowski problem. It turns out that in three of the four schemes one obtains similar behavior, only constrained evolution of the fixed lapse system leads to instabilities that presumably can be explained through an inappropriate choice of initial data.

In sections 6 and 7 we compare the free and the constrained symplectic schemes on the Schwarzschild space-time. Section 6 deals with Brown's generalized harmonic system, showing that the results of both schemes are similar. In section 7 the fixed lapse system is considered. There it turns out that errors which arise at the boundaries in the free evolution are suppressed in the constrained scheme.

Our numerical experiments thus confirm this remarkable property of the constrained scheme that enforces the momentum constraints. But they also show that constrained evolution is not beneficial when applied to a constrained Hamiltonian system that does not enforce the momentum constraints.

We finish the article by showing how the fixed lapse Hamiltonian can be derived in appendix A.

\section{Hamiltonian formulations of general relativity with dynamical shift and densitized lapse}

As motivated in the introduction we are interested in Hamiltonian formulations of general relativity that provide strongly hyperbolic equations of motion. There is no unique way to find such a formulation. We follow the ideas proposed by Brown \cite{brown-2008}. That is, we include hyperbolic drivers for the shift and the densitized lapse functions. Our approach differs in some details, and we summarize it in what follows.

The systems of interest are second order in space and first order in time. A definition of strong hyperbolicity for those systems is given e.g. in \cite{Gundlach:2005ta}. For strongly hyperbolic systems it can be proven that they possess a well-posed Cauchy problem \cite{Nagy:2004td,calabrese_1228487}, they are hence preferred in numerical applications.

In the construction we start from the well-known \emph{Super-Hamiltonian} \cite{Dirac-1958,Dirac-1959} (see also \cite{Anderson_York_4556,gourgoulhon-2007}). There the position variables are provided by the 3-metric $h_{ij}$ and the corresponding canonical momenta are denoted $\pi^{ij}$. They are related to the extrinsic curvature $K_{ij}$ through
$ \pi^{ij} = \sqrt{h}\left(K^l{}_l h^{ij} - K^{ij} \right)$.
Here $h$ is the determinant of the 3-metric $h_{ij}$.
The Super-Hamiltonian takes the form
\begin{equation}\label{super-H}
{\mathcal H}_S = \int (\alpha C + \beta^i C_i) \,d^3x
\end{equation}
with freely specifiable (external) functions $\alpha$ and $\beta^i$.
The {\it densitized lapse} $\alpha$ is related to the lapse function $N$ through $\alpha = N/\sqrt{h}$, and
$\beta^i$ is the {\it shift vector}. The functions $C$ and $C_i$ are given by
\begin{equation}
\label{ham-constr}
C = \pi^{ij}\pi_{ij}-\frac{1}{2}\pi^i{}_i\pi^j{}_j- h R\,,
\end{equation}
where $R$ is the Ricci scalar of the metric $h_{ij}$, and%
\footnote{Notice that $\pi^{jk}$ is a tensor density of weight $+1$.}
\begin{equation}
\label{mom-constr}
C_i = - 2 h_{ij} D_k\pi^{jk}.
\end{equation}
The equations $C=0$ and $C_i=0$ are the Hamiltonian and momentum constraints respectively.

\subsection{Shift and densitized lapse as dynamical variables}

According to \cite{brown-2008} we can extend the phase space and use a modified Hamiltonian such that shift and densitized lapse become dynamical variables. Here we treat the shift as a momentum and the densitized lapse as a position.%
\footnote{The reasons for this choice are discussed in section \ref{sec:constrained_integration_gen_harmonic} or in \cite{Richter:2008pr}.}
We introduce their canonically conjugate variables $\gamma_i$ and $\sigma$ respectively. The canonical symplectic two-form for the extended phase-space of $(h_{ij},\alpha,\gamma_i;\pi^{ij},\sigma,\beta^i)$ is then $dh_{ij}\wedge d\pi^{ij}+d\alpha\wedge d\sigma+d\gamma_{i}\wedge d\beta^{i}$.

The interpretation of $\beta^i$ as momentum variables may lead to complications if one wants to transform from the Hamiltonian equations of motion to an action with Euler-Lagrange equations. The 3+1 decomposition of the Einstein-Hilbert action explicitly depends on the shift. As an action it must not contain momentum variables, but only positions and their time derivatives. Yet, given a Hamiltonian it depends on its concrete form whether the shift can be expressed in terms of positions and their time derivatives. Hence, it may be impossible to perform the Legendre transformation to obtain an appropriate action.

Here we take the Hamiltonian formulation of the equations as fundamental and do not consider possible problems concerning the action and corresponding Euler-Lagrange equations further.

Yet, in \cite{brown-2008} Brown gets the constraints
\begin{align}
\label{eq:sigma_gamma_constraints}
 \sigma&=0&&\mbox{and}&\gamma_i &= 0
\end{align}
as primary constraints from the Einstein-Hilbert action. Since for us the Hamiltonian formulation is fundamental we are not limited to this case. However, of course we are interested in an Hamiltonian that provides dynamical equations for $h_{ij}$ and $\pi^{ij}$ that are consistent with the equations of motion in general relativity. We will see that if \eqref{eq:sigma_gamma_constraints} is satisfied then this requirement is automatically fulfilled with an appropriate extended Hamiltonian.

We define
\begin{align}
 \label{eq:Hamiltonian_gen_harmonic_general}
 \mathcal H_g := \int d^3x\left(\hat\Lambda\sigma+\hat\Omega^i\gamma_i\right),
\end{align}
where $\hat\Lambda$, $\hat\Omega^i$ are functions of the canonical variables that are linear in $\pi^{ij}$, $\sigma$, $\gamma_i$ and in the first spatial derivatives $\partial_k h_{ij}$, $\partial_i\alpha$, $\partial_i\beta^j$, with coefficients that depend on $h_{ij}$, $\alpha$ and $\beta^j$.

Due to the structure of the gauge Hamiltonian $\mathcal H_g$ the canonical Hamiltonian equations of motion for $h_{ij}$ and $\pi^{ij}$ that correspond to the full Hamiltonian $\mathcal H_S+\mathcal H_g$ are equivalent to the dynamical part of the ADM equations when the constraints \eqref{eq:sigma_gamma_constraints} are satisfied. But the freedom in $\mathcal H_g$ can be used to introduce quite general hyperbolic drivers for $\alpha$ and $\beta^i$.

For concrete calculations we need to make one particular choice for the functions $\hat\Lambda$ and $\hat\Omega^i$. The simulations that we discuss here are based on a generalized harmonic system. We set
\begin{align}
\label{eq:Lambda_gen_harmonic}
 \hat\Lambda &= \beta^i \partial_i\alpha
-\alpha \partial_i\beta^i
+\frac12\alpha^3\sigma
\end{align}
and
\begin{align}
\label{eq:Omega_gen_harmonic}
\hat\Omega^i &= -\beta^j\partial_j\beta^i
- h \alpha^2\Gamma^i_{jk}h^{jk}
+ h \alpha h^{ij}D_j\alpha
- \frac12 h \alpha^3h^{ij}\gamma_j.
\end{align}
There are also other possible choices, and in \cite{brown-2008} Brown discusses some of them.

{\allowdisplaybreaks
The canonical Hamiltonian equations of motion, i.e. the \emph{generalized harmonic system}, is now
\begin{subequations}
\label{eq:gen_harmonic_eq_of_motion}
\begin{align}
\label{eq:gen_harmonic_dot_hij}
\dot h_{ij} &= 2\alpha\pi_{ij}-\alpha \pi^k{}_k h_{ij} +D_i\beta_j +D_j\beta_i,\\
\nonumber
\label{eq:gen_harmonic_dot_piij}
\dot \pi^{ij} &=
\alpha \pi^k{}_k\pi^{ij}-2\alpha\pi^{i}{}_k\pi^{kj}+\alpha hh^{ij}R-\alpha h R^{ij}-\pi^{ik}D_k\beta^j
-\pi^{jk}D_k\beta^i\\
\nonumber
&\qquad
+hD^iD^j\alpha-hh^{ij}D^kD_k\alpha+hD_l(\beta^l\pi^{ij})\\
\nonumber
&\qquad
-\frac12\alpha^3 {h} \gamma^i \gamma^j
+\frac12\alpha^3 {h} h^{ij} \gamma_k \gamma^k
-\frac12\alpha {h} \gamma^j \partial^i\alpha
-\frac12\alpha {h} \gamma^i \partial^j\alpha\\
\nonumber
&\qquad
-\frac14\alpha^2 {h} h^{kl} \gamma^j \partial^ih_{kl}
-\frac14\alpha^2 {h} h^{kl}\gamma^i \partial^jh_{kl}
-\frac12\alpha^2 {h} \partial^i\gamma^j
-\frac12\alpha^2 {h} \partial^j\gamma^i\\
\nonumber
&\qquad
+\alpha {h} h^{ij} \gamma^k \partial_k\alpha
+\frac14\alpha^2 {h} h^{jk} h^{lm} \gamma^i \partial_kh_{lm}
+\frac14\alpha^2 {h} h^{ik} h^{lm} \gamma^j \partial_kh_{lm}\\
&\qquad
+\alpha^2 {h} h^{ij} h^{kl} \partial_k\gamma_l
-\frac12\alpha^2 {h} h^{il} h^{jm} \gamma^k \partial_lh_{km}
-\frac12\alpha^2 {h} h^{il} h^{jm} \gamma^k \partial_mh_{kl},\\
\label{eq:gen_harmonic_dot_alpha}
\dot \alpha &= \alpha^3\sigma + \beta^i \partial_i\alpha - \alpha\partial_i\beta^i,\\
\nonumber
\label{eq:gen_harmonic_dot_sigma}
\dot \sigma &= - C+\beta^i\partial_i\sigma+2\sigma\partial_i\beta^i
+\alpha h h^{il}\partial_i\gamma_l\\
&\qquad-\frac{3}{2}\left(\alpha^2\sigma^2-\alpha^2h\gamma_i\gamma_j h^{ij}\right)
-\alpha h h^{jk}h^{il}\gamma_l\partial_i h_{jk}+\alpha h \gamma_lh^{ij}h^{lk}\partial_i h_{jk},\\
\label{eq:gen_harmonic_dot_gammai}
\dot\gamma_i &= C_i + 2\sigma\partial_i\alpha+\alpha\partial_i\sigma-\gamma_j\partial_i\beta^j+\gamma_i\partial_j\beta^j+\beta^j\partial_j\gamma_i,\\
\label{eq:gen_harmonic_dot_betai}
\dot\beta^i &= \beta^j\partial_j\beta^i -\alpha^2h\Gamma^j_{jk}h^{ik}+\alpha^2h\Gamma^i_{jk}h^{jk}+\alpha^3hh^{ij}\gamma_j-\alpha hh^{ij}D_j\alpha.
\end{align}
\end{subequations}
We see that \eqref{eq:gen_harmonic_dot_hij} is also one of the ADM equations, i.e. the dynamical behavior of $h_{ij}$ is not changed. The equation \eqref{eq:gen_harmonic_dot_piij} for $\pi^{ij}$ differs from the corresponding ADM equation, it involves additional terms. But these terms vanish when $\gamma_i$ vanishes. Hence, in order to describe the dynamical behavior of $\pi^{ij}$ correctly we need to require that $\gamma_i=0$.
}

In \eqref{eq:sigma_gamma_constraints} we not only have the constraint $\gamma_i=0$, but also $\sigma=0$. Yet, the equations of motion for the ``physical'' variables $h_{ij}$ and $\pi^{ij}$ already agree with the ADM equations when only $\gamma_i=0$ is satisfied. One may hence conclude that $\sigma$ does not need to vanish.

However, for $\sigma\neq0$ we find instabilities already when $h_{ij}$ and $\pi^{ij}$ are exact solutions of the ADM equations. In particular, when $\gamma_i=0$ and the vector constraints are satisfied then equation \eqref{eq:gen_harmonic_dot_gammai} implies
\begin{align}
 0&=2\sigma\partial_i\alpha+\alpha\partial_i\sigma.
\end{align}
It follows that $\sigma(\bfx,t)=c(t)/\alpha^2(\bfx,t)$. Then, if the scalar constraint is satisfied as well we obtain from \eqref{eq:gen_harmonic_dot_alpha} and \eqref{eq:gen_harmonic_dot_sigma} that
\begin{align}
 c(t)=\frac{2c_0}{2-c_0(t-t_0)},\quad\mbox{where }c_0=c(t_0).
\end{align}
Hence, we obtain stable results only when $|c_0(t-t_0)|$ remains small. Therefore it is obvious that $\sigma=0$ should be satisfied, too. The full set of constraints to be satisfied is thus
\begin{align}
\label{eq:complete_constraints}
 C&=0,&
 C_i&=0,&
 \sigma&=0,&
 \gamma_i&=0.
\end{align}

\subsubsection{Constrained integration of the generalized harmonic system}
\label{sec:constrained_integration_gen_harmonic}

In \cite{Richter:2008pr} we proposed a numerical integration scheme for the ADM equations that enforces the momentum constraints in a holonomically constrained Hamiltonian system and turns the Hamilton constraint function from a weak to strong invariant.

The main steps to achieve this were to interpret the shift as a momentum and to impose the holonomic constraints $\gamma_i=0$ as well as an appropriate gauge condition. The obvious question is now whether an analogous scheme with similar properties can be constructed also for the generalized harmonic system \eqref{eq:gen_harmonic_eq_of_motion}.

Unfortunately this is not the case. For the ADM equations the momentum constraints could be enforced, because they were hidden constraints of $\gamma_i=0$ (i.e. $\dot\gamma_i = C_i$). Here we do not have this direct relation, but we have \eqref{eq:gen_harmonic_dot_gammai} instead. Hence, when we enforce $\gamma_i=0$ at each time then we only get
\begin{align}
0 &= C_i + 2\sigma\partial_i\alpha+\alpha\partial_i\sigma,
\end{align}
i.e. only this combination of the momentum constraints and $\sigma=0$ can be enforced. This might still be beneficial, but we cannot control the momentum constraints separately and we cannot expect that the Hamilton constraint function becomes a strong invariant.

Obviously the generalized harmonic system \eqref{eq:gen_harmonic_eq_of_motion} here behaves differently than the ADM equations because of the terms that contain $\alpha$ and $\sigma$. These terms will virtually always appear when the densitized lapse is treated as a dynamical variable. Therefore we investigate in the next section a Hamiltonian system where $\alpha$ is an external field, and only $\beta^i$ is subject to a dynamical gauge condition.

\subsection{Hyperbolic shift driver with fixed densitized lapse}
\label{sec:fixed_lapse_system}

Since we cannot enforce the momentum constraints in a holonomically constrained Hamiltonian system when the densitized lapse function is a dynamical variable, we are interested in a Hamiltonian formulation where $\alpha$ is an external field. Thus, we consider the phase space of $(h_{ij},\gamma_i;\pi^{ij},\beta^i)$ with symplectic two-form $dh_{ij}\wedge d\pi^{ij}+d\gamma_i\wedge d\beta^i$.

We define another gauge Hamiltonian
\begin{align}
\label{eq:H_beta_3+1}
 \mathcal H_\beta &= \int d^3x \hat\Omega^i\gamma_i,
\end{align}
and need to find a function $\hat\Omega^i$ such that the canonical Hamiltonian equations of motion that correspond to $\mathcal H_S+\mathcal H_\beta$ become strongly hyperbolic.

It turns out that this requirement is very restrictive when $\alpha$ is an external field. We are able to construct an appropriate function:
\begin{align}
\label{eq:Omega_i_alpha_ext}
 \hat\Omega^i &=
 -\beta^j\partial_j\beta^i
 -\frac47 \alpha^2 \Gamma^i_{jk} h h^{jk}
 +\frac67 \alpha^2 \Gamma^k_{kj} h h^{ji}\\
\nonumber&\qquad\qquad\qquad
 +\frac27 \alpha hh^{ij}D_j\alpha
 +\frac27 \alpha^3 h h^{ij}\gamma_j,
\end{align}
but every other possible choice for $\hat\Omega^i$ that we investigated leads to a system with the same principal part as \eqref{eq:alpha_ext_system} (see appendix \ref{app:other_Ham_formulations}).

{\allowdisplaybreaks
The canonical Hamiltonian equations of motion that we derive from $\mathcal H_S+\mathcal H_\beta$ using \eqref{eq:Omega_i_alpha_ext} are now
\begin{subequations}
\label{eq:alpha_ext_system}
\begin{align}
\label{eq:alpha_ext_system_hij}
\dot h_{ij} &= 2\alpha\pi_{ij}-\alpha \pi^k{}_k h_{ij} +D_i\beta_j +D_j\beta_i,\\
\label{eq:alpha_ext_system_piij}
\dot\pi^{ij} &=
 2\alpha\pi^{i}{}_k\pi^{jk} - \alpha\pi^{ij}\pi^{k}{}_k
 + \alpha h R^{ij} - \alpha h h^{ij}R\\
\nonumber
&\quad
 + \pi^{jk}D_k\beta^i
 + \pi^{ik}D_k\beta^j
 - \pi^{ij}D_k\beta^k
 - \beta^k D_k\pi^{ij}
 - h D^i D^j\alpha
 + h h^{ij}D^k D_k\alpha\\
\nonumber&\quad
 +  \frac\alpha7
\bigg(
  4\alpha h\Gamma^{k}_{lm} (h^{il} h^{jm}-h^{ij} h^{lm})\gamma_k
 - 3\alpha h\Gamma^l_{lm} (h^{im}\gamma^j + h^{jm}\gamma^i - 2h^{ij}\gamma^m)\\
\nonumber&\qquad\qquad
 - 2\alpha^2 h \gamma^i\gamma^j
 + 2\alpha^2 h h^{ij}\gamma^k\gamma_k
 + 2 \alpha h D^i\gamma^j
 + 2 \alpha h D^j\gamma^i
 - 6 \alpha h h^{ij}D_k\gamma^k\\
\nonumber&\qquad\qquad
 + 3 h \gamma^j D^i\alpha
 + 3 h \gamma^i D^j\alpha
 - 10 h h^{ij}\gamma_k D^k\alpha\bigg),\\
\label{eq:alpha_ext_system_gammai}
 \dot\gamma_i &= C_i - \gamma_j\partial_i\beta^j + \gamma_i\partial_j\beta^j + \beta^j\partial_j\gamma_i,\\
\label{eq:alpha_ext_system_betai}
 \dot\beta^i &= - \beta^j\partial_j\beta^i + \frac27\left(
   3 \alpha^2\Gamma^j_{jk} h h^{ik}
 - 2\alpha^2\Gamma^i_{jk}h h^{jk}
 + 2\alpha^3 h \gamma^i
 + \alpha h D^i\alpha\right).
\end{align}
\end{subequations}
In what follows we denote these equations the \emph{fixed lapse system}. Again \eqref{eq:alpha_ext_system_hij} is one of the ADM equations and \eqref{eq:alpha_ext_system_piij} becomes an ADM equation when $\gamma_i=0$.
}

We also see from \eqref{eq:alpha_ext_system_gammai} that the momentum constraints vanish when $\gamma_i$ vanishes identically. Therefore, in contrast to the generalized harmonic system \eqref{eq:gen_harmonic_eq_of_motion} we can enforce the momentum constraints in a holonomically constrained system. Since the dynamical behavior of $h_{ij}$ and $\pi^{ij}$ is then equivalent to the ADM equations, we also get that the Hamilton constraint function becomes a strong invariant \cite{Anderson_York_4556}:
\begin{align}
 (\partial_t-\beta^iD_i)C=0.
\end{align}
Even if this property does not extend to the space discretization, it is an extra bonus for this momentum-constrained formulation.

In the numerical examples we will use 1+1~dimensional simplifications of the two formulations \eqref{eq:gen_harmonic_eq_of_motion} and \eqref{eq:alpha_ext_system}.

\section{Discrete Hamiltonian and numerical integration methods}

To apply numerical methods one approximates the continuous functions $h_{ij}$, $\gamma_i$, $\alpha$, $\pi^{ij}$, $\beta^i$ and $\sigma$ through objects with finitely many degrees of freedom.%
\footnote{We use finite differences and piecewise constant functions, as described in section \ref{sec:finite_differences}.}
We collect the finite number of unknowns in four vectors $\bfq$, $\bfp$, $\bfgamma$ and $\bfbeta$. For the discretized system \eqref{eq:gen_harmonic_eq_of_motion} $\bfq$ and $\bfp$ correspond to the discretizations of $(h_{ij},\alpha)$ and $(\pi^{ij},\sigma)$ respectively, whereas for the system \eqref{eq:alpha_ext_system} these vectors correspond to the discretizations of $h_{ij}$ and $\pi^{ij}$ respectively. The ordering in $\bfq$, $\bfp$, $\bfgamma$ and $\bfbeta$ is chosen such that components corresponding to the same grid point are ordered consecutively.

For both considered systems the full Hamiltonian ($\mathcal H_S+\mathcal H_g$ respectively $\mathcal H_S+\mathcal H_\beta$) consists of terms that are either quadratic in the momentum variables $(\pi^{ij},\sigma,\beta^i)$ or independent of them. Any reasonable discretization of the full Hamiltonian will hence assume the form%
\footnote{For the fixed lapse system \eqref{eq:alpha_ext_system} we ignore the dependence on the discrete densitized lapse in the notation.}
\begin{align}
 \label{H-disc}
 H (\bfq,\bfp) = \frac12 \bfp^T \bfS(\bfq) \bfp + 
U(\bfq,\bfgamma) + \bfbeta^T \bfD(\bfq) \bfp + \frac12\bfbeta^T \bfE(\bfq,\bfgamma) \bfbeta,
\end{align}
where $\bfS(\bfq)$, $\bfD(\bfq)$ and $\bfE(\bfq,\bfgamma)$ are matrices of the appropriate dimensions.

The canonical equations of motion for this discrete Hamiltonian are
\begin{subequations}
\label{eq:general_Ham_system}
\begin{align}
\label{eq:general_Ham_system_q}
 \dot\bfq &= \bfS(\bfq)\bfp+\bfD(\bfq)^T\bfbeta,\\
\label{eq:general_Ham_system_gamma}
 \dot\bfgamma &= \bfE(\bfq,\bfgamma)\bfbeta+\bfD(\bfq)\bfp,\\
\label{eq:general_Ham_system_p}
 \dot\bfp &= -\frac12\bfp^T\bfnabla_\bfq\bfS(\bfq)\bfp
  -\bfnabla_\bfq U(\bfq,\bfgamma) -\bfbeta^T\bfnabla_\bfq\bfD(\bfq)\bfp
  -\frac12\bfbeta^T\bfnabla_\bfq\bfE(\bfq,\bfgamma)\bfbeta,\\
\label{eq:general_Ham_system_beta}
 \dot\bfbeta &= -\frac12\bfbeta^T\bfnabla_\bfgamma\bfE(\bfq,\bfgamma)\bfbeta
  -\bfnabla_\bfgamma U(\bfq,\bfgamma).
\end{align}
\end{subequations}

In section \ref{sec:SV-method} we introduce a free symplectic method that integrates \eqref{eq:general_Ham_system} numerically and in section \ref{sec:RATTLE-method} a constrained scheme is presented. The latter additionally requires the following constraints
\begin{align}
\label{eq:discrete_constraints}
 \bfgamma &=\bfzero,&
 \bfg(\bfq) &= \bfzero.
\end{align}
The equation $\bfg(\bfq)=\bfzero$ is a gauge condition and fixes $\bfbeta$. It must be chosen such that the following matrix is invertible (see also \cite{Richter:2008pr,HaLW06})
\begin{align}
 \bfA(\bfq) := \bfnabla_\bfq\bfg(\bfq)^T\bfD(\bfq)^T.
\end{align}
Time differentiation of $\bfg(\bfq)=\bfzero$ and using \eqref{eq:general_Ham_system_q} for $\dot\bfq$ gives
\begin{align}
 \bfnabla_\bfq\bfg(\bfq)^T\bfS(\bfq)\bfp+\bfA(\bfq)\bfbeta &= \bfzero,
\end{align}
which shows that indeed $\bfbeta$ is determined by the gauge condition, when $\bfA(\bfq)$ is invertible.

A candidate for the choice of the function $\bfg$ is a discretization of the \emph{Dirac gauge}, $\partial_j(h^{1/3} h^{ij})=0$ \cite{Dirac-1959,Bonazzola_PhysRevD.70.104007}. With this choice, $\bfA(\bfq)$ is a discretized second-order elliptic differential operator. In section \ref{sec:finite_differences} we discuss other gauge conditions with a similar structure for a simplified system.

\subsection{The St\"ormer-Verlet method}
\label{sec:SV-method}

A standard symplectic integrator for Hamiltonian systems is the St\"ormer--Verlet scheme (see, e.g., \cite{HaLW03}). When applied to
\eqref{eq:general_Ham_system}, a step from values $(\bfq^n,\bfgamma^n,\bfp^n,\bfbeta^n)$ at time $t^n$ to values 
$(\bfq^{n+1},\bfgamma^{n+1},\bfp^{n+1},\bfbeta^{n+1})$ at time $t^{n+1}=t^n+\dt$ reads as follows:
{\allowdisplaybreaks
\begin{eqnarray}
\label{sv-p1}
\bfp^{n+1/2} &=& \bfp^n - \frac\dt 2
\Big(
 \frac12 (\bfp^{n+1/2})^T \bfnabla_\bfq \bfS(\bfq^{n})\bfp^{n+1/2}
 +\bfnabla_\bfq U(\bfq^{n},\bfgamma^{n})
\\
\nonumber
&&  \qquad\qquad\qquad + \:
 (\bfbeta^{n+1/2})^T\bfnabla_\bfq \bfD(\bfq^{n})\bfp^{n+1/2}
\\
\nonumber
&&  \qquad\qquad\qquad + \:
 \frac12(\bfbeta^{n+1/2})^T\bfnabla_\bfq \bfE(\bfq^{n},\bfgamma^{n})\bfbeta^{n+1/2}
\Big)
\\
\label{sv-beta1}
\bfbeta^{n+1/2} &=& \bfbeta^n - \frac\dt 2
\Big(
 \bfnabla_\bfgamma U(\bfq^{n},\bfgamma^{n})
+ \frac12(\bfbeta^{n+1/2})^T\bfnabla_\bfgamma \bfE(\bfq^{n},\bfgamma^{n})\bfbeta^{n+1/2}
\Big)
\\
\label{sv-q}
\bfq^{n+1} &=& \bfq^n + \frac\dt 2 
\Big(
\bigl(\bfS(\bfq^{n+1}) + \bfS(\bfq^n)\bigr)\bfp^{n+1/2}
\\
\nonumber
&&  \qquad\qquad\quad +\: \bigl(\bfD(\bfq^{n+1})^T + \bfD(\bfq^n)^T\bigr)\bfbeta^{n+1/2}
\Big)
\\
\label{sv-gamma}
\bfgamma^{n+1} &=& \bfgamma^n + \frac\dt 2 
\Big(
\bigl(\bfE(\bfq^{n+1},\bfgamma^{n+1}) + \bfE(\bfq^n,\bfgamma^n)\bigr)\bfbeta^{n+1/2}
\\
\nonumber
&&  \qquad\qquad\quad +\: \bigl(\bfD(\bfq^{n+1}) + \bfD(\bfq^n)\bigr)\bfp^{n+1/2}
\Big)
\\
\label{sv-p2}
\nonumber
\bfp^{n+1} &=& \bfp^{n+1/2} - \frac\dt 2
\Big(
 \frac12 (\bfp^{n+1/2})^T \bfnabla_\bfq \bfS(\bfq^{n+1})\bfp^{n+1/2}
 +\bfnabla_\bfq U(\bfq^{n+1},\bfgamma^{n+1})
\\
&&  \qquad\qquad\qquad + \:
 (\bfbeta^{n+1/2})^T\bfnabla_\bfq \bfD(\bfq^{n+1})\bfp^{n+1/2}
\\
\nonumber
&&  \qquad\qquad\qquad + \:
 \frac12(\bfbeta^{n+1/2})^T\bfnabla_\bfq \bfE(\bfq^{n+1},\bfgamma^{n+1})\bfbeta^{n+1/2}
\Big)
\\
\label{sv-beta2}
\bfbeta^{n+1} &=& \bfbeta^{n+1/2} - \frac\dt 2
\Big(
 \frac12 (\bfnabla_\bfgamma U(\bfq^{n+1},\bfgamma^{n+1})
\\
\nonumber
&&  \qquad\qquad\qquad + \:
\frac12(\bfbeta^{n+1/2})^T\bfnabla_\bfgamma \bfE(\bfq^{n+1},\bfgamma^{n+1})\bfbeta^{n+1/2}
\Big).
\end{eqnarray}
}

In this system the computationally most expensive terms are subsumed in the expressions $\bfnabla_\bfq U$ and $\bfnabla_\bfgamma U$. There is only one evaluation per step of those terms.

The system \eqref{sv-p1}-\eqref{sv-beta2} can be decomposed into three substeps,
the first two, (\eqref{sv-p1},\eqref{sv-beta1}) and (\eqref{sv-q},\eqref{sv-gamma}) being implicit in $(\bfp^{n+1/2},\bfbeta^{n+1/2})$ and $(\bfq^{n+1},\bfgamma^{n+1})$, respectively. They are solved by fixed-point iteration, which is local at every grid point. The last substep (\eqref{sv-p2},\eqref{sv-beta2}) is explicit.

With that method the constraints are not explicitly taken into account. Yet, since in the continuous equations of motion the constraints are satisfied for all times as long as they are satisfied for the initial data, one may expect that the constraint violations in the discrete case remain small.

\subsection{The RATTLE method}
\label{sec:RATTLE-method}
The RATTLE method (\cite{hc_andersen}, \cite[Section VII.1]{HaLW06},
\cite[Chapter 7]{LeR04}) is an extension of the St\"ormer--Verlet method to holonomically constrained systems. It is symplectic and time-reversible, of second order accuracy, and enforces both the holonomic and the derived hidden constraints in the numerical solution. When applied to (\ref{eq:general_Ham_system}),(\ref{eq:discrete_constraints}), a step of the RATTLE method consists of the following equations, which form a nonlinear system for $\bfq^{n+1}$, $\bfgamma^{n+1}$, $\bfp^{n+1}$ and $\bfbeta^{n+1}$.
\begin{enumerate}
\item First half-step for the momentum variables:
{\allowdisplaybreaks
\begin{subequations}
\label{eq:first_half_step}
\begin{eqnarray}
\label{p-half}
\nonumber
\bfp^{n+1/2} &=& \bfp^n - \frac\dt 2
\Big(
 \frac12 (\bfp^{n+1/2})^T \bfnabla_\bfq \bfS(\bfq^{n})\bfp^{n+1/2}
 + \bfnabla_\bfq U(\bfq^{n},\bfgamma^n)
\\
&&  \qquad\qquad + \:
 (\bfbeta^{n+1/2})^T\bfnabla_\bfq \bfE(\bfq^{n},\bfgamma^{n})\bfbeta^{n+1/2}
\\
\nonumber
&&  \qquad\qquad + \:
 (\bfbeta^{n+1/2})^T\bfnabla_\bfq \bfD(\bfq^{n})\bfp^{n+1/2}
+ \bfnabla_\bfq \bfg(\bfq^{n})\bflambda^{n,+}
\Big)
\\
\label{beta-half}
\nonumber
\bfbeta^{n+1/2} &=& \bfbeta^n - \frac\dt 2
\Big(
(\bfbeta^{n+1/2})^T\bfnabla_\bfgamma \bfE(\bfq^{n},\bfgamma^{n})\bfbeta^{n+1/2}
\\
&&  \qquad\qquad + \:
\bfnabla_\bfgamma U(\bfq^{n},\bfgamma^n)
+ \bfmu^{n,+}
\Big)
\end{eqnarray}
\end{subequations}
\item Full step for the position variables:
\begin{subequations}
\label{eq:full_step}
\begin{eqnarray}
\label{q-new}
\bfq^{n+1} &=& \bfq^n + \frac\dt 2 
\Big(
\bigl(\bfS(\bfq^{n+1}) + \bfS(\bfq^n)\bigr)\bfp^{n+1/2}
\\
\nonumber
&&  \qquad\qquad\quad +\: \bigl(\bfD(\bfq^{n+1})^T + \bfD(\bfq^n)^T\bigr) \bfbeta^{n+1/2}
\Big)
\\
\label{gamma-new}
\bfgamma^{n+1} &=& \bfgamma^n + \frac\dt 2 
\Big(
\bigl(\bfE(\bfq^{n+1},\bfgamma^{n+1}) + \bfE(\bfq^n,\bfgamma^n)\bigr)\bfbeta^{n+1/2}
\\
\nonumber
&&  \qquad\qquad\quad +\: \bigl(\bfD(\bfq^{n+1}) + \bfD(\bfq^n)\bigr) \bfp^{n+1/2}
\Big)
\qquad (\bfgamma^n=\bfzero)
\end{eqnarray}
\end{subequations}
\item Position constraints:
\begin{subequations}
\label{eq:pos_constraints}
\begin{eqnarray}
\label{q-constr}
\bfg(\bfq^{n+1}) &=& \bfzero
\\
\label{gamma-constr}
\bfgamma^{n+1} &=& \bfzero
\end{eqnarray}
\end{subequations}
\item Second half-step for the momentum variables:
\begin{subequations}
\label{eq:second_half_step}
\begin{eqnarray}
\label{p-half-2}
\nonumber
\bfp^{n+1} &=& \bfp^{n+1/2}
\\
\nonumber&&
 - \frac\dt 2
\Big(
 \frac12 (\bfp^{n+1/2})^T \bfnabla_\bfq \bfS(\bfq^{n+1})\bfp^{n+1/2}
 + \bfnabla_\bfq U(\bfq^{n+1},\bfgamma^{n+1})
\\
&&  \qquad + \:
 (\bfbeta^{n+1/2})^T\bfnabla_\bfq \bfE(\bfq^{n+1},\bfgamma^{n+1})\bfbeta^{n+1/2}
\\
\nonumber
&&  \qquad + \:
 (\bfbeta^{n+1/2})^T\bfnabla_\bfq \bfD(\bfq^{n+1})\bfp^{n+1/2}
+ \bfnabla_\bfq \bfg(\bfq^{n+1})\bflambda^{n+1,-}
\Big)
\\
\label{beta-half-2}
\nonumber
\bfbeta^{n+1} &=& \bfbeta^{n+1/2} - \frac\dt 2
\Big(
(\bfbeta^{n+1/2})^T\bfnabla_\bfgamma \bfE(\bfq^{n+1},\bfgamma^{n+1})\bfbeta^{n+1/2}
\\
&&  \qquad\qquad + \:
\bfnabla_\bfgamma U(\bfq^{n+1},\bfgamma^{n+1})
+ \bfmu^{n+1,-}
\Big)
\end{eqnarray}
\end{subequations}
\item Momentum constraints:
\begin{subequations}
\label{eq:mom_constraints}
\begin{eqnarray}
\label{beta-constr}
\bfnabla_\bfq \bfg(\bfq^{n+1})^T 
\bfS(\bfq^{n+1})\bfp^{n+1} + 
\bfA(\bfq^{n+1})\bfbeta^{n+1} &=& \bfzero
\\[2mm]
\label{p-constr}
\bfD(\bfq^{n+1})\bfp^{n+1}&=&\bfzero\,.
\end{eqnarray}
\end{subequations}
}
\end{enumerate}
In the last equation \eqref{p-constr} we used the fact that for our Hamiltonian the matrix $\bfE(\bfq,\bfgamma)$ vanishes when $\bfgamma=\bfzero$.

Equations (\ref{eq:first_half_step})--(\ref{eq:pos_constraints}) determine
$\bfq^{n+1}$, and the system (\ref{eq:second_half_step}),(\ref{eq:mom_constraints}) determines $\bfp^{n+1}$. The equations can be solved by an iterative procedure that requires only the solution of
linear systems with the matrices $\bfA(\bfq^n)$ and $\bfA(\bfq^{n+1})$ and their transposes. This procedure is described in \cite{Richter:2008pr}, and we do not discuss it here.

\section{1+1 dimensional test cases}
\label{sect:11}
In this article we are interested in the numerical properties of the discretized systems \eqref{eq:gen_harmonic_eq_of_motion} and \eqref{eq:alpha_ext_system}. Our numerical experiments are based on a simplified model that arises in certain highly symmetric space-times. Essentially we assume that the 3-metric is diagonal, depends only on the coordinate $x^1$ and has just two independent components, $h_{33}=\zeta h_{22}$. The details are described in \cite{Richter:2008pr}.

We distinguish between two essentially 1+1 dimensional classes of solutions of Einstein's equations, namely the spherically symmetric space-times where $\zeta:=\sin^2 x^2$ and a second class where $\zeta\equiv 1$. The latter includes a perturbed Minkowski geometry.

It is then natural to define
\begin{align}
 \tilde h &:= \frac12\left(h_{22}+\zeta^{-1} h_{33}\right),&
 \tilde \pi &:= \pi^{22}+\zeta\pi^{33},
\end{align}
and it turns out that $\tilde\pi$ is indeed the canonical momentum corresponding to $\tilde h$.

In the spherically symmetric case we now consider the equatorial hypersurface, i.e., $x^2=\pi/2$ and $\zeta=1$.
Using the new variables $\tilde h$ and $\tilde\pi$ one can derive 1+1 dimensional counterparts for the Hamiltonians $\mathcal H_S$, $\mathcal H_g$ and $\mathcal H_\beta$. The ``1+1-dimensional Super-Hamiltonian'' becomes
\begin{align}
\label{eq:super_Hamiltonian1+1}
\nonumber
 \mathcal H_S^1&=\int dx\bigg[
    \alpha\left(
    \frac12\pi^{11}\pi^{11}h_{11}h_{11}-
    \pi^{11}\tilde\pi h_{11}\tilde h\right)\\
    \nonumber&\qquad\qquad
    -\alpha\left(
    \frac12 \partial \tilde h\partial \tilde h-
    2 \tilde h \partial^2 \tilde h+
    \tilde h\partial \tilde h\partial \log(h_{11})+2\xi h_{11}\tilde h\right)\\
    &\qquad\qquad + 2\pi^{11}h_{11}\partial\beta+
    \pi^{11}\beta\partial h_{11}+
                \tilde\pi\beta\partial \tilde h
    \bigg],
\end{align}
and the 1+1-dimensional gauge Hamiltonians are
\begin{align}
\label{eq:gen_harmonic_Hamiltonian1+1}
 \mathcal H_g^1&=\int dx
 \bigg[2\alpha^2\tilde h\gamma\partial\tilde h
 + \alpha \tilde h^2 \gamma \partial\alpha
 - \frac12\alpha^3\tilde h^2\gamma^2
 - \beta\gamma\partial\beta\\
\nonumber &\qquad\qquad\qquad
 + \beta\sigma\partial\alpha
 - \alpha\sigma\partial\beta
 + \frac12\alpha^3\sigma^2\bigg],\\
\label{eq:alpha_ext_Hamiltonian1+1}
 \mathcal H_\beta^1&= \int dx\bigg[
   \frac{12}7 \alpha^2\tilde h \gamma\partial\tilde h
 + \frac27 \alpha^2\tilde h^2 \gamma\partial\log h_{11}
 + \frac27 \alpha \tilde h^2 \gamma \partial\alpha \\
\nonumber&\qquad\qquad\qquad
 - \frac27\alpha^3\tilde h^2\gamma^2
 -\beta\gamma\partial\beta\bigg].
\end{align}
Here we have $\xi=1$ in the spherically symmetric case and $\xi=0$ if $\zeta\equiv 1$.

These Hamiltonians have a similar structure as their 3+1-dimensional counterparts \eqref{super-H} and \eqref{eq:Hamiltonian_gen_harmonic_general},\eqref{eq:H_beta_3+1}, their discretization will therefore be of the form \eqref{H-disc}.

\subsection{Space Discretization in the 1+1 dimensional setting}
\label{sec:finite_differences}

The spatial discretization of the Hamiltonians \eqref{eq:super_Hamiltonian1+1}, \eqref{eq:gen_harmonic_Hamiltonian1+1} and \eqref{eq:alpha_ext_Hamiltonian1+1} is done similarly as in \cite{Richter:2008pr}. We use piecewise constants functions to approximate $h_{11}$, $\tilde h$, $\alpha$, $\gamma$, $\pi^{11}$, $\tilde\pi$, $\sigma$ and $\beta$, where $\beta$ and $\gamma$ are discretized on a staggered grid. Yet, for the discrete spatial derivatives we use a modified approach.

We distinguish between four types of terms. The first three types contain only one (first or second) derivative, they are discretizations of terms of the form $f^1f^2\partial f^3$ or $f^1f^2\partial^2 f^3$, with functions $f^1$, $f^2$ and $f^3$. The fourth type contains a product of first derivatives, it is the discrete counterpart of $f^1\partial f^2\partial f^3$.
\begin{enumerate}
 \item In the first type of terms the discrete counterpart of the functions $f^i$ are defined on the same grid. An example is the term $2\alpha\tilde h\partial^2\tilde h$ in \eqref{eq:super_Hamiltonian1+1}. For these terms we approximate the derivative with centered second order finite differences as in \cite{Richter:2008pr}:
\begin{align}
\label{eq:discrete_deriv}
\nonumber
 \partial f^3(x_i) &\rightarrow (D_{0}f^3)_i = (f^3_{i+1}-f^3_{i-1})/2\Delta x,\\
 \partial^2 f^3(x_i) &\rightarrow (D_{+}D_{-}f^3)_i = (f^3_{i+1}-2f^3_i+f^3_{i-1})/\Delta x^2.
\end{align}
\item In the second type of terms the discrete function $f^3$ is defined on the staggered grid and $f^1$, $f^2$ on the non staggered grid. An example is the term $2\pi^{11}h_{11}\partial\beta$ in \eqref{eq:super_Hamiltonian1+1}. In this term we discretize the derivative using one-sided finite differences
\begin{align}
 (f^1f^2\partial f^3)(x_i) \rightarrow f^1_if^2_i(D_{-}f^3)_i = f^1_if^2_i(f^3_i-f^3_{i-1})/\Delta x.
\end{align}
Since $f^3$ is defined on the staggered grid this formula is second order accurate at the grid points of the non staggered grid.
\item For the third type of terms the discrete functions $f^1$ and $f^3$ are defined on the non staggered grid and $f^2$ on the staggered grid. An example is the term $\pi^{11}\beta\partial h_{11}$ in \eqref{eq:super_Hamiltonian1+1}. For this term we average the function $f^1$ and use one-sided finite differences for the derivative of $f^3$:
\begin{align}
\nonumber
 (f^1f^2\partial f^3)(x_i) \rightarrow \left(f^1_i+f^1_{i+1}\right)&f^2_i(D_{+}f^3)_i/2 =\\
&= \left(f^1_i+f^1_{i+1}\right)f^2_i(f^3_{i+1}-f^3_{i})/2\Delta x.
\end{align}
Again, since $f^2$ is defined on the staggered grid this formula is second order accurate at the grid points of the staggered grid.
\item Finally, the fourth type of terms deals with discretizations of $f^1\partial f^2\partial f^3$, where the functions $f^i$ are defined on the same grid. An example is the term $\alpha\partial\tilde h\partial\tilde h$ in \eqref{eq:super_Hamiltonian1+1}. There we use a combination of one-sided finite differences
\begin{align}
\label{eq:discrete_deriv_square}
 (f^1\partial f^2\partial f^3)(x_i) \rightarrow f^1_i((D_{-}f^2)_i(D_{-}f^3)_i+(D_{+}f^2)_i(D_{+}f^3)_i)/2.
\end{align}
Again this formula is second order accurate.
\end{enumerate}

There are two reasons to change the discretization of spatial derivatives. At first, in \cite{Richter:2008pr} we observed instabilities with periodic boundary conditions and an even number of grid points.

We can explain these instabilities as follows. When we discretize using centered finite differences, i.e. $f^1\partial f^2\partial f^3(x_i)\rightarrow f^1_i (D_{0}f^2)_i (D_{0}f^3)_i$, then the derivative of the resulting Hamiltonian with respect to $f_i^3$ contains a term $f^1_i(D_{0}^2 f^2)_i$ (see also e.g. \cite{calabrese_1228487} for the definition of $D_{0}$, $D_{+}$ and $D_{-}$). This is a centered finite difference formula with an extended stencil ($D_0^2$ discretization).
If we linearize the corresponding equations and check the well-posedness of the result using the techniques proposed in \cite{calabrese_1228487} we find instabilities when the number of grid points is even. This is not the case when we discretize the spatial derivatives using \eqref{eq:discrete_deriv_square}.

Moreover, when the discretizations \eqref{eq:discrete_deriv} and \eqref{eq:discrete_deriv_square} are used then we can (for periodic boundary conditions) perform partial integrations also in the discrete Hamiltonian, because
\begin{align}
 \frac12\sum_{i}& f_i \left((D_{+}g)_i (D_{+}h)_i + (D_{-}g)_i (D_{-}h)_i\right)=\\
\nonumber
 &= -\sum_{i} f_i g_i (D_{+}D_{-}h)_i
    -\frac12\sum_{i} g_i \left((D_{+}f)_i (D_{+}h)_i + (D_{-}f)_i (D_{-}h)_i\right).
\end{align}

\paragraph{Boundary conditions and discrete gauge conditions.}
Concerning the treatment of boundary conditions we proceed analogously as in \cite{Richter:2008pr}. We apply periodic boundary conditions for the perturbed Minkowski problem and Dirichlet boundary conditions in the Schwarzschild space-time. The discrete boundary conditions are treated using ghost zones.

As in \cite{Richter:2008pr} the gauge conditions for the RATTLE scheme are derived from continuous gauge conditions of the form
\begin{align}
 g(h_{11},\tilde h,\alpha)=0.
\end{align}

When we are interested in the generalized harmonic system (corresponding to $\mathcal H_S^1+\mathcal H_g^1$) then we choose
\begin{align}
\label{eq:rad_harm_Sch_gauge}
 g(h_{11},\tilde h,\alpha) = \partial\left(h_{11}\tilde h^{-2}\exp(2M\xi(R-1))\right),
\end{align}
where $\xi=1$ in the spherically symmetric case and $\xi=0$ when $\zeta\equiv 1$.
When we perform calculations with the fixed lapse system (corresponding to $\mathcal H_S^1+\mathcal H_\beta^1$) then we set
\begin{align}
\label{eq:alpha_ext_Sch_gauge}
 g(h_{11},\tilde h,\alpha) = \partial\left(\alpha h_{11}\tilde h^6\right).
\end{align}
The reason for this choice is clearly that the analytic solutions for which we test the schemes (see section \ref{sec:test_scenarios}) satisfy those gauge conditions.

For the generalized harmonic system in the spherically symmetric case ($\xi=1$) we see that the gauge condition depends on the mass parameter $M$. Hence, if we did not know the analytical solutions then it was not possible to apply this gauge condition. However, in principle one can also apply other gauge conditions, e.g. Dirac gauge \cite{Dirac-1959,Bonazzola_PhysRevD.70.104007}, in the RATTLE method, but the results are then not compareable to the results of the free evolution scheme, because one needs to use different initial data (those that satisfy the gauge condition).

\subsection{Test scenarios}
\label{sec:test_scenarios}

In sections \ref{sec:ex_Mink}--\ref{sec:ex_Sch_alpha_ext} we perform numerical experiments with the St\"ormer-Verlet and the RATTLE method. There we check how well the numerical schemes reproduce a perturbed Minkowski space-time and the Schwarzschild space-time in two different coordinate systems. The analytical solutions are the following.

\subsubsection{A perturbed Minkowski metric}
\label{sec:Minkowski}

The Minkowski metric describes a flat space-time, the analytical solution is (with $x=x^1$, $y=x^2$, $z=x^3$)
\begin{align}
 ds^2=-dt^2+dx^2+dy^2+dz^2.
\end{align}
It is easy to check that for $t=\,$const. slicing this solution really is in the class of solutions with $\zeta\equiv 1$ and thus the numerical schemes we described are applicable.

Here we perturb the Minkowski initial data such that
\begin{align}
 h_{11} &= 1+\varepsilon_{11},&
 \tilde h&=1+\tilde\varepsilon,&
 \alpha&=1+\varepsilon_\alpha,&
 \gamma&=\varepsilon_\gamma
\end{align}
and
\begin{align}
 \pi^{11}&=\delta^{11},&
 \tilde\pi&=\tilde\delta,&
 \sigma&=\delta_\sigma,&
 \beta&=\delta_\beta.
\end{align}
In the numerical examples we will choose the perturbations $\varepsilon$, $\delta$ to be Gaussian functions of width $1/20$ and height $10^{-6}$.
To avoid problems with boundaries we apply periodic boundary conditions.

\subsubsection{Schwarzschild space-time}
\label{sec:Schwarzschild}
In the numerical experiments that deal with the Schwarzschild space-time we consider two different coordinate systems. They are chosen such that we get solutions of the equations of motion for the Hamiltonians $\mathcal H_S^1+\mathcal H_g^1$ and $\mathcal H_S^1+\mathcal H_\beta^1$ respectively. We denote these coordinate systems ``radially harmonic'' and ``fixed lapse coordinates'' respectively.

\paragraph{Radially harmonic coordinates.}
From the Schwarzschild space-time in standard coordinates $(t,r,\theta,\phi)$ we come to the radially harmonic coordinates via the coordinate transformation
\begin{align}
 R(r)&=1+\frac1{2M}\log\left(\frac{r}{r-2M}\right),&
 r(R)&=\frac{2M}{1-\exp(2M(1-R))}.
\end{align}
That is, $R\in(1,\infty)$ and $R=\infty$ is the horizon, whereas $R=1$ is spatial infinity.

The 4-metric in the new coordinate system is
\begin{align}
 ds^2&=-e^{2M(1-R)}dt^2+e^{2M(1-R)}r^4 dR^2+r^2 d\Omega^2,
\end{align}
where $d\Omega^2$ is the volume element of the sphere and $r=r(R)$. From the 4-metric we read off the following expressions
\begin{align}
\nonumber
 h_{11} &= r^4 e^{2M(1-R)},&
 \tilde h &= r^2,&
 \alpha &= r^{-4},\\
 \pi^{11}&=0,&
 \tilde\pi&=0,&
 \beta&=0.
\end{align}
This solution satisfies the canonical Hamiltonian equations of motion that correspond to $\mathcal H_S^1+\mathcal H_g^1$ and the gauge condition \eqref{eq:rad_harm_Sch_gauge} for $\xi=1$.

\paragraph{Fixed lapse coordinates.}
The fixed lapse coordinates for the Schwarzschild space-time are obtained from the standard Schwarzschild coordinates $(t,r,\theta,\phi)$ via the coordinate transformation
\begin{align}
 R(r)&=\left(\frac{r}{2M}\right)^{11},&
 r(R)&=2MR^{1/11}.
\end{align}

The 4-metric in the new coordinate system is
\begin{align}
 ds^2&=\left(1-R^{-1/11}\right)dt^2+\frac{4M^2}{121 R^{20/11}\left(1-R^{-1/11}\right)}dR^2+4M^2 R^{2/11} d\Omega^2
\end{align}
and we get
\begin{align}
 \nonumber
 h_{11} &= \frac{4M^2}{121 R^{20/11}\left(1-R^{-1/11}\right)},&
 \pi^{11}&=0,\\
 \tilde h &= 4M^2R^{2/11},&
 \tilde\pi&=0,\\
 \nonumber
 \alpha &= 11(2M)^{-3}R^{8/11}\left(1-R^{-1/11}\right),&
 \beta&=0.
\end{align}
This solution satisfies the canonical Hamiltonian equations of motion that correspond to $\mathcal H_S^1+\mathcal H_\beta^1$ and the gauge condition \eqref{eq:alpha_ext_Sch_gauge} for $\xi=1$.

\section{A perturbed Minkowski problem: comparison of the four schemes}
\label{sec:ex_Mink}

In this section we compare the results of the four numerical schemes that we obtain when we combine the 1+1 dimensional generalized harmonic system (corresponding to $\mathcal H_S^1+\mathcal H_g^1$) and the fixed lapse system (corresponding to $\mathcal H_S^1+\mathcal H_\beta^1$) with the St\"ormer-Verlet and the RATTLE integrator.

\paragraph{Simulation data.}
We apply the schemes to the perturbed Minkowski problem described in section \ref{sec:Minkowski}. The perturbations are Gaussian functions of width $1/20$ and height $\epsilon=10^{-6}$.

When we apply the RATTLE scheme then we set $\varepsilon_\gamma = 0$. The justification for this choice is that in the constrained scheme the equation $\bfgamma=\bfzero$ is enforced for each time step and for the initial data it is always easy to satisfy it.

Moreover, we set $\varepsilon_\alpha=0$ for the fixed lapse system, because there $\alpha$ is an external field and we assume that the correct function is given in advance.

In the simulations we use 50 grid points and the size of the time step is a quarter of the spatial grid spacing, $\Delta t=\Delta x/4$.

\begin{figure}
 \begin{tabular}{cc}
  \includegraphics[scale=0.65]{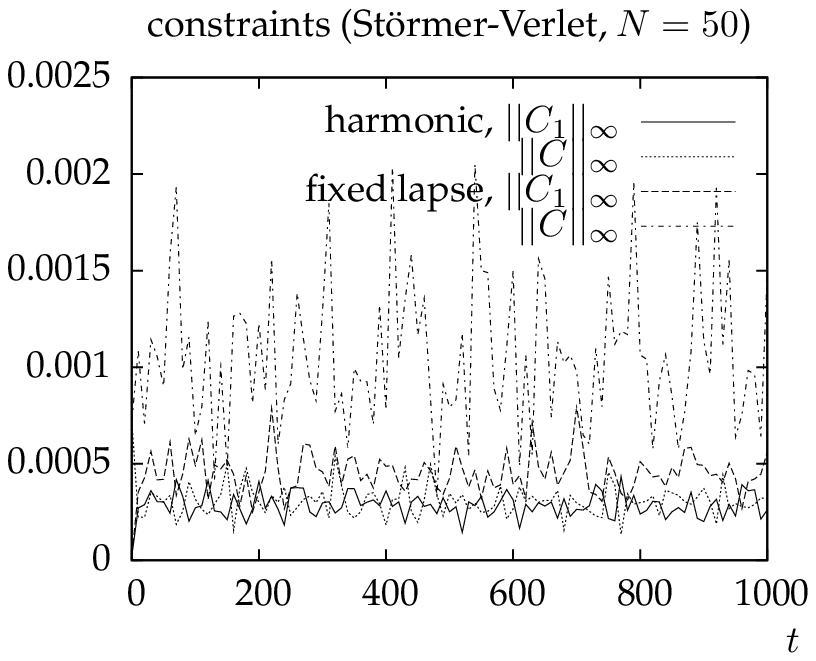}&
  \includegraphics[scale=0.65]{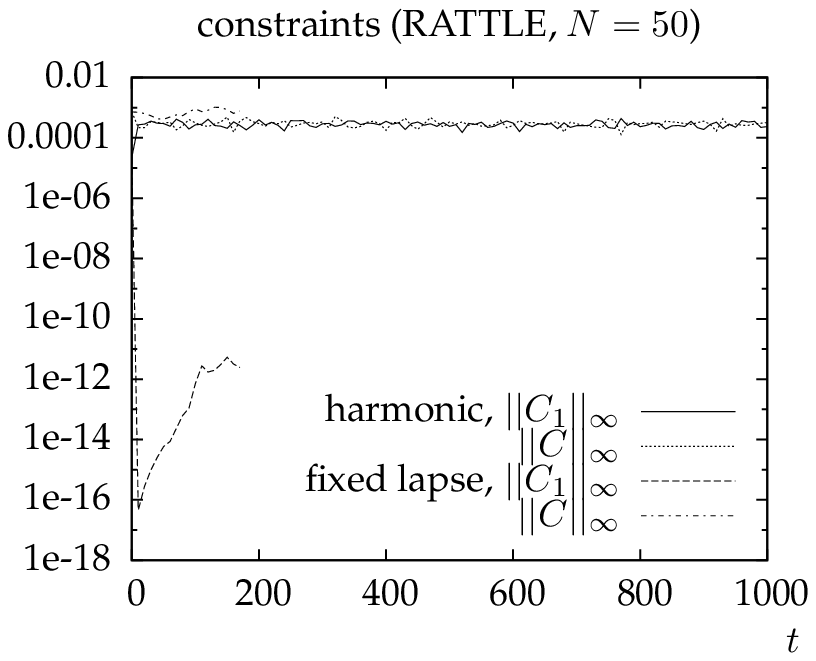}
 \end{tabular}
 \caption{Perturbed Minkowski space-time: The maximum norm of Hamilton and momentum constraints, $C$ and $C_1$ respectively. Left: Results of the St\"ormer-Verlet method. Right: Results of the RATTLE method.
}
 \label{fig:constraints_Minkowski}
\end{figure}

\paragraph{Simulation results.}
We see in figure \ref{fig:constraints_Minkowski} that the results for both systems of equations are similar when the free integration scheme using St\"ormer-Verlet is applied. Moreover the RATTLE scheme applied to the 1+1 dimensional generalized harmonic system also provides similar results.

In those cases the evolution is stable at least until $t=1000$ and we do not see growing constraints or errors. In particular the Hamilton and momentum constraint functions stay at almost the same size of about $2\cdot 10^{-4}$--$2\cdot 10^{-3}$. However, these functions do not vanish.

The RATTLE scheme applied to the 1+1 dimensional fixed lapse system behaves differently. By construction it is clear that the momentum constraints become small. They are the hidden constraints of $\bfgamma=\bfzero$ and in our implementation of the RATTLE scheme they are enforced with an accuracy of at least $10^{-10}$. However, at some point the evolution cannot be continued in this scheme, because the algorithm to solve the nonlinear system does not converge within $10^{4}$ iteration steps. The problem there is to satisfy $\bfgamma = \bfzero$, i.e. \eqref{gamma-constr}.

\paragraph{Discussion of results.}
The results that we get for this example indicate that strong hyperbolicity of the equations of motion is still essential when we use symplectic integrators for Hamiltonian systems.

In \cite{Richter:2008pr} we discussed tests of the same problem. There we used discretized ADM equation in the evolution and got rapidly growing high frequency errors. This is not the case here, and at least for three schemes all other desireable properties that the symplectic integration of the ADM equations showed (in particular the conservation of the harmonic energies) are still valid. This behavior was expected in the beginning, because the initial value problem of strongly hyperbolic systems is well-posed whereas it may be ill-posed for systems that are only weakly hyperbolic (the ADM equations are weakly hyperbolic but not strongly hyperbolic) \cite{Gundlach:2005ta,calabrese_1228487}.

Now, out of the four schemes that we investigated there are three that provide the expected results.
From the fourth scheme, the constrained integration of the fixed lapse system, we see that to obtain favorable results it is not sufficient to start from a strongly hyperbolic system.

From a computational point of view we find that the reason for the problems in that scheme is basically that the momentum constraints cannot be enforced with the required accuracy, because the algorithm to solve the nonlinear system (\ref{eq:first_half_step})--(\ref{eq:pos_constraints}) does not converge.

The solution of this scheme itself points to the interpretation that singularities occur during the evolution. In particular we observe large gradients in $\tilde\pi$ and $\beta$. It is evident that these problems are not purely numerical, since we get quite similar results when we take a smaller time step, $\Delta t=\Delta x/8$ and also with higher resolutions, $N=100$ and $N=200$.

It is not clear whether this is a problem of the coordinate system or of the underlying solution. The algorithm that we apply projects onto the (momentum) constrained hypersurface in each time step, and in particular in the first one. As the constraints are quite large for the considered initial data the result of this projection can be that the problem is no longer Minkowski with a small perturbation, but a solution that develops a physical singularity after some time. However, when we start from initial data with smaller constraints, e.g. by taking the height of the Gaussian perturbations as $\epsilon=10^{-7}$, we can evolve to later times.

Since the fourth scheme provides better results when we start from initial data with small constraints one may expect that constrained integration is feasible when the initial data are chosen appropriately. This is supported by the calculations presented in section \ref{sec:ex_Sch_alpha_ext} where we find that it can be even beneficial to use this scheme.

\section{Schwarzschild space-time: free vs. constrained generalized harmonic evolution}
\label{sec:ex_Sch_gen_harmonic}

Here we discuss the St\"ormer-Verlet and the RATTLE scheme for the 1+1 dimensional generalized harmonic system (corresponding to $\mathcal H_S^1+\mathcal H_g^1$). We apply these schemes to the Schwarzschild space-time in radially harmonic coordinates (see section \ref{sec:Schwarzschild}). The mass parameter is chosen to be $M=1$ and the boundaries are at $R_l=2$ and $R_r=3$.

\begin{figure}
 \begin{tabular}{cc}
  \includegraphics[scale=0.65]{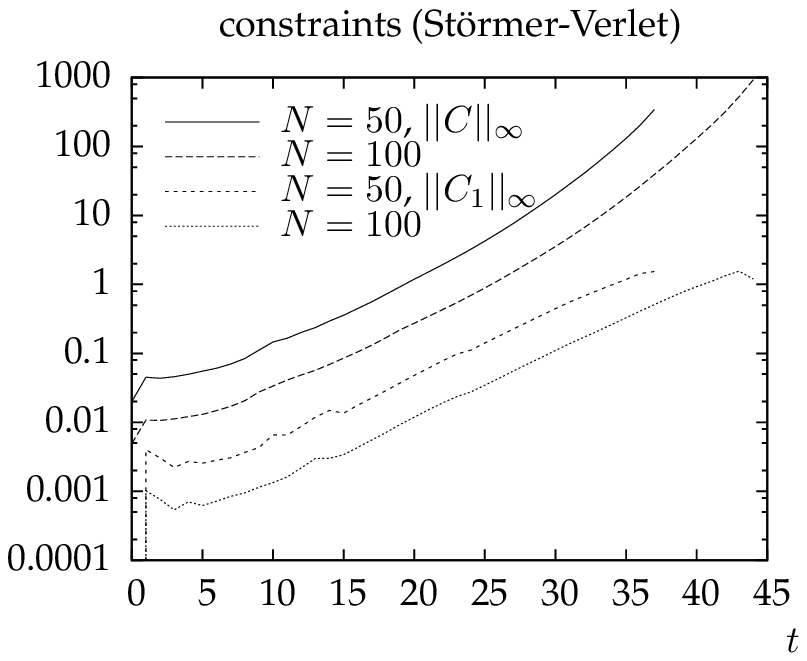}&
  \includegraphics[scale=0.65]{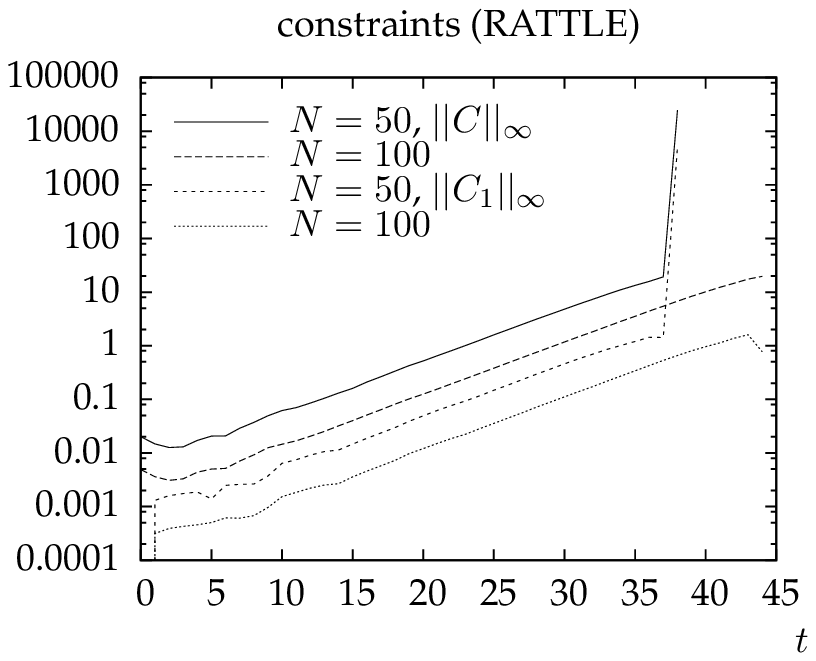}
 \end{tabular}
 \caption{Schwarzschild space-time: The maximum norm of the Hamilton and momentum constraints, $C$ and $C_1$ respectively, in the generalized harmonic evolution system. Left: Results of the free evolution scheme. Right: Results of the constrained evolution scheme.}
 \label{fig:Schwarzschild_harm_cnstr}
\end{figure}

\paragraph{Simulation results.}
In figure \ref{fig:Schwarzschild_harm_cnstr} we see that for both, free and constrained evolution the Hamilton and momentum constraints are growing exponentially. The same behavior can be observed for the error of the functions themselves. At some point the evolution breaks down.

This happens almost at the same time for both integration methods. But one can evolve to later times if one uses a finer grid. Moreover, with better resolution errors become quadratically smaller.

\paragraph{Discussion of results.}

For the generalized harmonic system in the Schwarz\-schild space-time the results of both schemes are qualitatively similar. This is in contrast to \cite{Richter:2008pr} where we observed that for the Schwarzschild space-time constrained symplectic evolution of ADM equations leads to better results than free evolution.

Yet, the difference there was that in the constrained evolution we were able to enforce the momentum constraints. Here we can only enforce a combination of the momentum constraints and $\sigma\equiv 0$. Hence, the boundary condition can lead to violations of the momentum constraints.

Again the results here are better than with the ADM evolution presented in \cite{Richter:2008pr}. There we found that the evolution breaks down at $t\approx 20$ for the constrained scheme and at $t\approx 3$ in the free evolution. Here we can evolve until $t\approx 40$ with both schemes. Moreover, in contrast to the free evolution using the ADM equations, errors are smaller in a fine grid and we can evolve longer than in a coarse grid.

\section{Schwarzschild space-time: free vs. constrained evolution with fixed densitized lapse}
\label{sec:ex_Sch_alpha_ext}

Finally we compare the St\"ormer-Verlet and the RATTLE scheme for the fixed lapse system (corresponding to $\mathcal H_S^1+\mathcal H_\beta^1$). We apply these schemes to the Schwarzschild space-time in fixed lapse coordinates (see section \ref{sec:Schwarzschild}). Again the mass parameter is chosen to be $M=1$ and the boundaries are at $R_l=2$ and $R_r=3$.

\begin{figure}
 \begin{tabular}{cc}
  \includegraphics[scale=0.65]{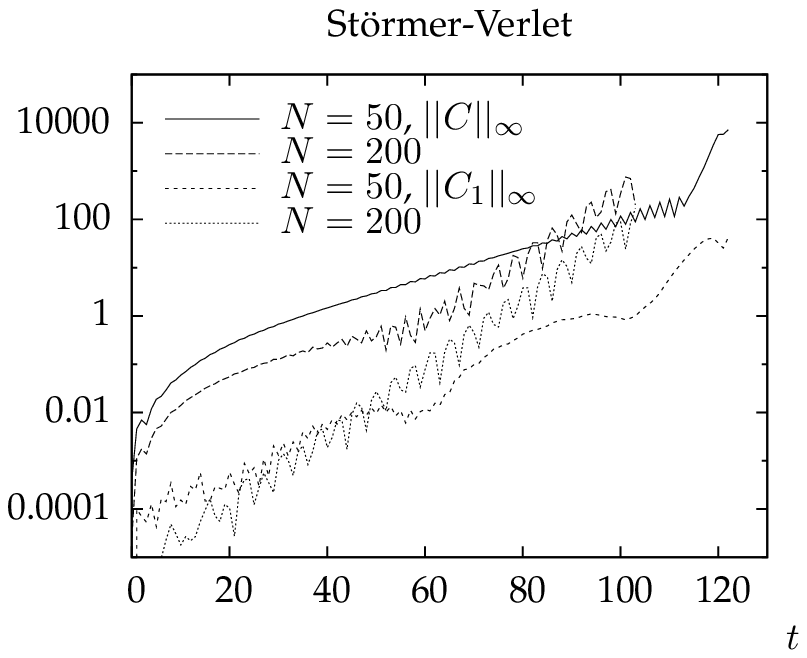}&
  \includegraphics[scale=0.65]{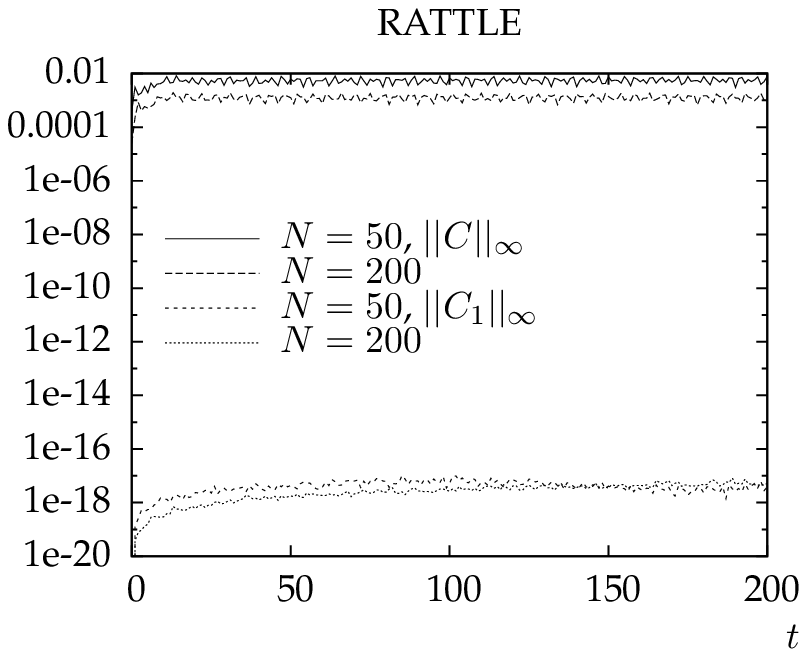}
 \end{tabular}
 \caption{Schwarzschild space-time: The maximum norm of the Hamilton and momentum constraints, $C$ and $C_1$ respectively, in the fixed lapse evolution system. Left: Results of the free evolution scheme. Right: Results of the constrained evolution scheme.}
 \label{fig:Schwarzschild_alpha_ext_cnstr}
\end{figure}

\begin{figure}
 \begin{tabular}{cc}
  \includegraphics[scale=0.65]{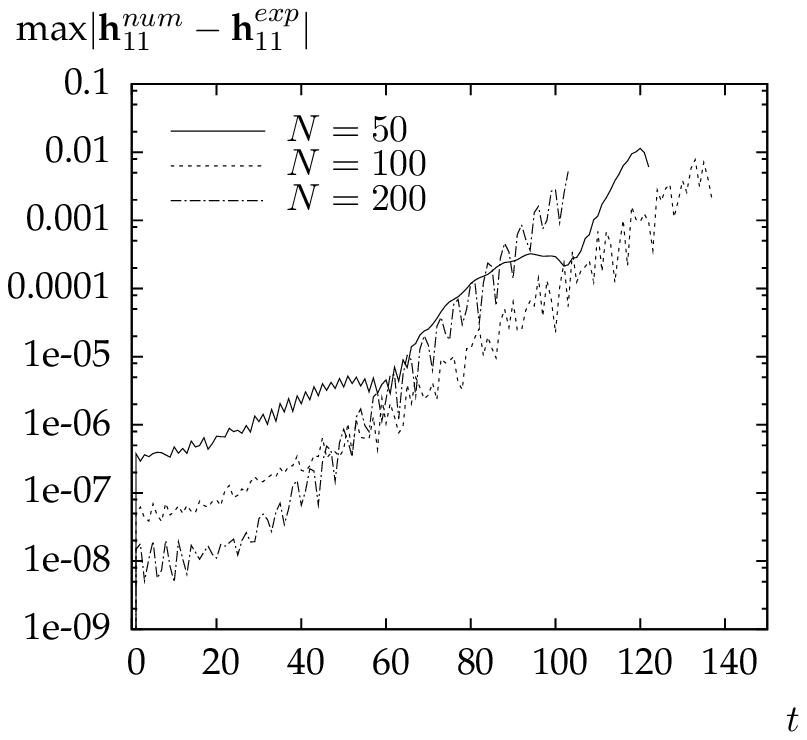}&
  \includegraphics[scale=0.65]{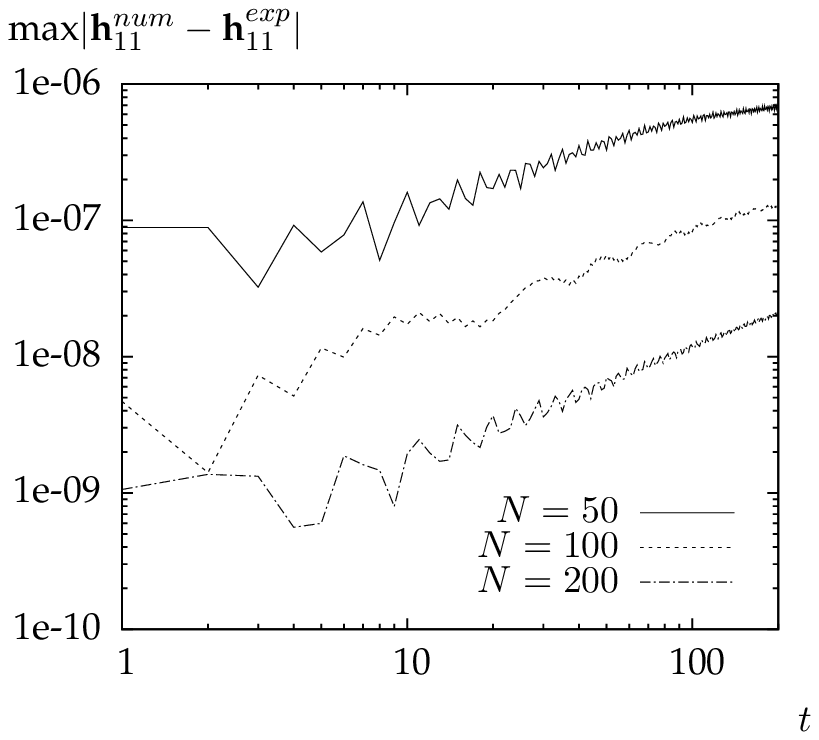}
 \end{tabular}
 \caption{Schwarzschild space-time: The maximum of the error of $h_{11}$, in the fixed lapse evolution system. Left: Results of the free evolution scheme. Right: Results of the constrained evolution scheme.}
 \label{fig:Schwarzschild_alpha_ext_error_h11}
\end{figure}

\paragraph{Simulation results.}
We see from figures \ref{fig:Schwarzschild_alpha_ext_cnstr} and \ref{fig:Schwarzschild_alpha_ext_error_h11} that with the free evolution scheme we obtain in the beginning (until $t\approx 40$) similar results as with the generalized harmonic system (see section \ref{sec:ex_Sch_gen_harmonic}). In particular constraints and errors are growing exponentially and become quadratically smaller when the resolution is increased.

At later times errors are still growing exponentially, but it may happen that they are bigger when a better resolution is used and that we can evolve to later times in a coarser grid.

In the constrained scheme we find that at least until $t=200$ the Hamilton constraint stays almost at the same size. For the errors of the functions themselves one finds at most linear growth. The momentum constraints are of course very small by construction.

\paragraph{Discussion of results.}
Here, as in \cite{Richter:2008pr} where we investigate the ADM system, we find better results when we apply the constrained RATTLE scheme.

In the free evolution we observe exponentially growing errors and instabilities. Presumably the reason for these problems are the naive Dirichlet boundary conditions that lead to constraint violations.

But in the RATTLE scheme there is no sign of growing constraints and moreover the errors of the functions themselves increase very slowly. Thus, for the systems and examples investigated so far constrained symplectic evolution of systems with boundaries leads to better results when the momentum constraints are enforced. It will be interesting to see whether this observation can be made in higher dimension, too.

In 1+1 dimensions, as there are no gravitational waves possible, we only need to worry about the boundary conditions to respect the constraints. In the constrained RATTLE scheme this is only partially ensured, it is in particular surprising that there are no increasing violations of the Hamilton constraints.

\section{Conclusion}
This article deals with symplectic numerical integration of strongly hyperbolic Hamiltonian formulations of general relativity. We discussed two appropriate Hamiltonians that are constructed by introducing hyperbolic drivers for the shift and in one system also for the densitized lapse. These are the fixed lapse system and a generalized harmonic system derived by Brown \cite{brown-2008}, respectively.

For both systems we performed numerical experiments using a free (St\"ormer-Verlet) and a constrained symplectic integrator (RATTLE). The free evolution schemes lead to similar results for both systems. In contrast to the ADM evolution \cite{Richter:2008pr} we do not see rapidly growing high frequency errors.
Also the good propagation properties of nearly conserved quantities that were found for some situations in the ADM evolution are recovered here. We thus conclude that it is indeed better to apply free symplectic integrators with strongly hyperbolic Hamiltonian systems.

With the constrained schemes the situation is different. When we apply the RATTLE method with the generalized harmonic system then we always got results that are very similar to the free evolution schemes. Since constrained integration requires much more computational resources than free evolution we thus conclude that constrained integration is not practicable with the generalized harmonic system.

Concerning constrained integration with the fixed lapse system we have shown one example where we observe instabilities in the evolution, but also one example where this scheme is much more stable than the others.

In the former case we argued that the problems occur due to constraint violations in the initial data. At the moment we cannot be certain that this is really the case, but the results can be interpreted in that way. We move this question to future investigations.

Yet, if the constraint violations in the initial data are indeed the reason for the problems then this RATTLE scheme is an interesting option to deal with the constraints for situations with appropriate initial data. This statement is supported by the results of that scheme for the fixed lapse and the ADM system (see \cite{Richter:2008pr}) with spherically symmetric solutions. There we observe stabilizing effects in the constrained evolution schemes. Since we do not find these effects when we apply the RATTLE method with the generalized harmonic system we suppose that our constrained evolution scheme has advantages when the momentum constraints are enforced in the equations of motion and the initial data satisfy the constraints.%
\footnote{The momentum constraints are enforced in the RATTLE scheme for the ADM equations and the fixed lapse system, but not for the generalized harmonic system.}
It will be interesting to see whether this conjecture can be supported by analytical studies of the underlying elliptic-hyperbolic system of equations or by further numerical investigations.

Another question to address in the future is the gauge choice. The systems that we consider here are quite restrictive for that matter. In particular in the fixed lapse system the requirement of strong hyperbolicity already fixed the Hamiltonian completely. This forced us to use special coordinates in the test examples, a factor that is clearly not satisfactory. One question is thus whether appropriate Hamiltonians exist that allow e.g. general Bona-Mass\'o slicings.

We also want to start to investigate examples in higher dimensions. We discussed that the free symplectic integrators are implicit, but that this implicitness is very mild such that symplectic integration does not need more computational resources than standard explicit methods. However, symplectic integrators should be applied to Hamiltonian systems. We expect that the right hand sides of discrete systems with that property will be more complicated than non Hamiltonian ones. But the size of the difference is not yet known.

\section*{Acknowledgment}
I would like to thank C. Lubich, J. M. Mart\'in-Garc\'ia and D. Brown for helpful suggestions and hints. This work was supported by the DFG grant SFB/Trans\-regio 7 ``Gravitational Wave Astronomy''.

\appendix

\section{Derivation of the fixed lapse Hamiltonian}
\label{app:other_Ham_formulations}

In this article we dealt with two strongly hyperbolic Hamiltonian formulations of general relativity, namely the generalized harmonic system \eqref{eq:gen_harmonic_eq_of_motion} and the fixed lapse system \eqref{eq:alpha_ext_system}. The former is similar to the generalized harmonic system derived in \cite{brown-2008}. Here we discuss the derivation of the latter.

As explained in section \ref{sec:fixed_lapse_system}, for this system the densitized lapse is an external field and the shift becomes a momentum variable. The canonical symplectic two-form for the phase-space of $(h_{ij},\gamma_i;\pi^{ij},\beta^i)$ is then $dh_{ij}\wedge d\pi^{ij}+d\gamma_{i}\wedge d\beta^{i}$.

Furthermore we assume the following form of the Hamiltonian
\begin{align}
 \label{eq:Hamiltonian_derive_fixed_lapse}
 \mathcal H = \mathcal H_S + \int d^3x\,\hat\Omega^i\gamma_i,
\end{align}
with a function $\hat\Omega^i$ that needs to be chosen appropriately.

To obtain this function we naturally start from the ansatz that Brown proposed in \cite{brown-2008}. Yet, here we are interested in the principal part of the resulting equations of motion only and omit all terms that do not contribute there.%
\footnote{In \cite{brown-2008} the terms that do not contribute to the principal part are needed in order to ensure that $\hat\Omega^i$ becomes a covariant vector under spatial diffeomorphisms and a scalar density of weight +2 under time reparametrizations.}
Hence, we start with
\begin{align}
\label{eq:Omega_general}
\hat\Omega^i &= -\beta^j\partial_j\beta^i
- C_2 h \alpha^2\Gamma^i_{jk}h^{jk}
- C_3 h \alpha^2\Gamma^j_{jk}h^{ik}\\
\nonumber&\qquad
+ C_5 h \alpha h^{ij}D_j\alpha
- C_6 h \alpha^3h^{ij}\gamma_j,
\end{align}
where $C_2$, $C_3$, $C_5$ and $C_6$ are constant parameters.

Having this ansatz for $\hat\Omega^i$ we can derive the equations of motion that correspond to the Hamiltonian \eqref{eq:Hamiltonian_derive_fixed_lapse} and their principal part. We obtain
\begin{align}
\nonumber
\left(
 \begin{array}{c}
  \dot h_{ij}\\
  \dot \beta^i\\
  \dot \pi^{ij}\\
  \dot \gamma_i
 \end{array}
\right)
 &\cong
\left(
\begin{array}{cccc}
 A^{klm}_{ij}\partial_m & A^\prime{}^{m}_{ijk}\partial_m & C_{ijkl} & C^\prime{}^k_{ij}\\
 \tilde A^{iklm}\partial_m & \bar A^{im}_k\partial_m & \tilde C^i_{kl} & \bar C^{ik}\\
 D^{ijklmn}\partial_m\partial_n & D^\prime{}^{ijmn}_{k}\partial_m\partial_n & G^{ijm}_{kl}\partial_m & G^\prime{}^{ijkm}\partial_m\\
 \tilde D^{klmn}_{i}\partial_m\partial_n & \bar D^{mn}_{ki}\partial_m\partial_n & \tilde G^m_{ikl}\partial_m & \bar G^{km}_{i}\partial_m
\end{array}
\right)
\left(
 \begin{array}{c}
  h_{kl}\\
  \beta^k\\
  \pi^{kl}\\
  \gamma_k
 \end{array}
\right)
\end{align}
where the symbol $\cong$ is used to denote equality up to lower order terms and for the non vanishing coefficients in this principal part we get
{\allowdisplaybreaks
\begin{align}
\nonumber
A^{klm}_{ij} &= \delta^k_i\delta^l_j\beta^m,\\
\nonumber
A^\prime{}^{m}_{ijk} &= \delta ^m_i h_{jk} +\delta ^m_j h_{ik},\\
\nonumber
\tilde A^{iklm} &= C_2 \alpha^2 h h^{ik} h^{lm}
                 - \frac12\left(C_2 - C_3 + C_5\right) \alpha^2 h h^{im} h^{kl},\\
\nonumber
\bar A^{im}_k &= \delta^i_k\beta^m,\\
\nonumber
C_{ijkl} &= 2\alpha h_{ik}h_{jl}-\alpha h_{ij}h_{kl},\\
\nonumber
\bar C^{ik} &= 2 C_6 \alpha^3 h h^{ik},\\
\nonumber
D^{ijklmn} &= \frac12\alpha h\big(h^{ij}(2 h^{ln}h^{km} - 3 h^{mn}h^{kl}),\\
\nonumber
\nonumber &\qquad
- h^{kn} h^{im} h^{jl}
- h^{ln} h^{ik} h^{jm}
+ h^{mn} h^{ik} h^{jl}
+ 2 h^{kl} h^{im} h^{jn}\big)
,\\
\nonumber
G^{ijm}_{kl} &= \delta^i_k\delta^j_l\beta^m,\\
\nonumber
G^\prime{}^{ijkm} &= -\frac12 \alpha^2 h
     \left(C_2 (h^{im} h^{jk} + h^{jm} h^{ik}) - (C_2 - C_3 + C_5) h^{ij} h^{km}\right),\\
\nonumber
\tilde G^m_{ikl} &= -2h_{ik}\delta^m_l,\\
\bar G^{km}_{i} &=\delta^k_i\beta^m.
\end{align}
}
The remaining coefficients $C^\prime$, $\tilde C$, $D^\prime$, $\tilde D$ and $\bar D$ vanish.

Now, according to \cite{Gundlach:2005ta} this second order in space evolution system is strongly hyperbolic if the second order principal symbol
\begin{align}
\label{eq:principal_symbol_fixed_lapse}
 P&=
\left(
\begin{array}{cccc}
 A^{klm}_{ij}n_m & A^\prime{}^{m}_{ijk}n_m & C_{ijkl} & C^\prime{}^k_{ij}\\
 \tilde A^{iklm}n_m & \bar A^{im}_kn_m & \tilde C^i_{kl} & \bar C^{ik}\\
 D^{ijklmn}n_mn_n & D^\prime{}^{ijmn}_{k}n_mn_n & G^{ijm}_{kl}n_m & G^\prime{}^{ijkm}n_m\\
 \tilde D^{klmn}_{i}n_mn_n & \bar D^{mn}_{ki}n_mn_n & \tilde G^m_{ikl}n_m & \bar G^{km}_{i}n_m
\end{array}
\right).
\end{align}
has real eigenvalues for all unit vectors $n^i$ and is diagonalizable in a regular way. Therefore the aim is now to choose the parameters $C_2,\ldots,C_6$ such that $P$ becomes diagonalizable.

The difference in comparison to the analysis of other formulations, like e.g. NOR \cite{Nagy:2004td}, is that we do not get strong hyperbolicity for a whole range of parameters. Instead we can derive nonlinear relations that the parameters need to satisfy. That is, only on a lower dimensional subset of the parameter space we obtain appropriate formulations. It is then very helpful if one has an algorithm to derive these relations. We apply the following one.

\subsection{Diagonalizability of a parameterized matrix}
\label{sec:algorithm}

Starting from a matrix $\mathcal P$ that depends on parameters $C_i$ we want to choose the parameters such that $\mathcal P$ is diagonalizable. The idea to find appropriate parameters is to adapt the algorithm for the calculation of the Jordan decomposition of a matrix.

The first step is to calculate the eigenvalues of $\mathcal P$, i.e. to find the roots of $\det(\mathcal P-\lambda\, \mathbb I)$. When $\mathcal P$ has a general form already this step might fail, because the analytical expressions for the eigenvalues are needed and for quasilinear systems the components of $\mathcal P$ depend on the fields. Moreover, the degree of the polynomial $\det(\mathcal P-\lambda\,\mathbb I)$ can become big and one obtains long expressions soon.
However, for our problem we indeed get analytical expressions for the eigenvalues.

Now, the multiplicity of the eigenvalues of $\mathcal P$ can be one or bigger than one. For the former eigenvalues nothing needs to be done, but for the latter it may happen that there are less eigenvectors than the multiplicity of the eigenvalue. In this case we proceed as follows.

Let $\lambda_1$ be an eigenvalue of multiplicity two or more. We calculate the kernels $K_1=ker(\mathcal P-\lambda_1\,\mathbb I)$ and $K_2=ker(\mathcal P-\lambda_1\,\mathbb I)^2$. If $K_1=K_2$ then the geometric multiplicity of $\lambda_1$ is the same as its algebraic multiplicity and nothing needs to be done. But if there are vectors in $K_2$ that are not in $K_1$ then the set of eigenvectors of $\lambda_1$ is not complete. That is, we need to restrict the possible choices of the parameters $C_i$.

To derive the corresponding equations we pick some vector $x\in K_2\setminus K_1$ and calculate $y=(\mathcal P-\lambda_1\, \mathbb I)x$. The vector $y$ depends on the parameters and we probably can choose them such that $y=0$. This leads to the desired relations. Of course these relations must not depend on the fields. It may happen that we cannot choose the parameters appropriately. In this case we don't get a strongly hyperbolic formulation.

Finally we use the derived relations to reduce the number of parameters in $\mathcal P$ and eventually repeat the procedure if there are still eigenvalues for which the geometric multiplicity is smaller than the algebraic one.

\subsection{The fixed lapse system}

Now, to analyse the diagonalizability of $P$ (defined in \eqref{eq:principal_symbol_fixed_lapse}) with the algorithm described in the previous section \ref{sec:algorithm} it is helpful to introduce an orthonormal basis $\{n^i,v^i,w^i\}$. When we expand the tensor indices in $h_{ij}$, $\pi^{ij}$, $\beta^i$ and $\gamma_i$ into this basis, denoting e.g. $h_{vv}=h_{ij}v^iv^j$, then $P$ decomposes into three blocks.

The first block, $P_1$, corresponds to the subsystem of the components $h_{vw}$ and $\pi^{vw}$. Its eigenvalues are $\beta^n \pm\alpha\sqrt{h}$ and it has a complete set of eigenvectors for each choice of the parameters $C_2,\ldots,C_6$.

The second block, $P_2$, comes from the subsystem of the components $h_{nn}$, $h_{vv}$, $h_{ww}$, $\pi^{nn}$, $\pi^{vv}$, $\pi^{ww}$, $\beta^{n}$ and $\gamma_{n}$. Its eigenvalues are $\beta^n\pm\alpha\sqrt{h}$ with multiplicity one and $\beta^n$, $\beta^n\pm\alpha\sqrt{h(1+C_2+C_3-C_5)}$ with multiplicity two. For a general choice of the parameters it turns out that $P_2$ is not diagonalizable.

We first consider the eigenvalue $\lambda_1=\beta^n$. If we apply the algorithm described in section \ref{sec:algorithm} then we find that diagonalizability of $P_2$ implies
\begin{align}
\label{eq:C6rel}
 C_6 &= -\frac18\left(9C_2^2+10C_2C_3+C_3^2-10 C_2C_5-2C_3C_5 +C_5^2\right).
\end{align}
Then, replacing $C_6$ in $P_2$ using \eqref{eq:C6rel} and applying the same steps as before for $\lambda_2=\beta^n-\alpha\sqrt{h(1+C_2+C_3-C_5)}$ we get
\begin{align}
\label{eq:C2rel}
 C_2 &= \frac19\left(4-C_3+C_5\right).
\end{align}

Here we assume that the parameters are chosen such that there are no additional degeneracies of eigenvalues. In particular we must choose $C_2+C_3-C_5\neq -1$ and $C_2+C_3-C_5\neq 0$. Furthermore, the reality condition on the eigenvalues implies $C_2+C_3-C_5\geq -1$.
Now, choosing $C_2$ and $C_6$ in this way $P_2$ is diagonalizable with real eigenvalues.

What remains is to analyse the third block of $P$, namely $P_3$. It comes from the subsystem of $h_{nv}$, $h_{nw}$, $\pi^{nv}$, $\pi^{nw}$, $\beta^v$, $\beta^w$, $\gamma_v$ and $\gamma_w$.%
\footnote{The first and third block do not appear for the simplified Hamiltonian $\mathcal H_S^1+\mathcal H_\beta^1$, because the corresponding system is just 1+1 dimensional.}
When we use \eqref{eq:C6rel} and \eqref{eq:C2rel} to remove $C_2$ and $C_6$ from $P_3$ then we obtain the eigenvalues $\beta^n\pm 1/3\alpha\sqrt{h(4-C_3+C_5)}$, both with multiplicity four. The algorithm of section \ref{sec:algorithm} applied with any of these eigenvalues then leads to
\begin{align}
 C_3 &= -\frac87+C_5.
\end{align}

Hence, altogether we get
\begin{align}
 C_2 &= \frac47,&
 C_3 &= -\frac87+C_5,&
 C_6 &= \frac27.
\end{align}
If these relations are satisfied then the eigenvalues of $P$ are automatically real. Hence, we have a strongly hyperbolic system and for $C_5=2/7$ we obtain the fixed lapse system \eqref{eq:alpha_ext_system}.

Since $P$ only depends on the difference $C_3-C_5$ and not on the sum of these parameters it is clear that other strongly hyperbolic formulations that are derived from the ansatz \eqref{eq:Omega_general} have the same principal part. There might be an exception when the parameters are chosen such that some eigenvalues coincide (e.g. when $C_2+C_3-C_5=0$). However, for our purposes it was sufficient to find a single strongly hyperbolic Hamiltonian system, and here we were able to derive one.

The calculations described above were performed using \emph{Mathematica} \cite{Mathematica} and the open-source package \emph{xTensor} for abstract tensor calculations developed by J. M. Mart\'in-Garc\'ia \cite{xAct}.

\end{document}